\documentstyle[twocolumn,aps,epsf]{revtex}
\begin{document}
\draft
\title{
How to bosonize fermions with non-linear energy dispersion}
\author{Peter Kopietz}
\address{
Institut f\"{u}r Theoretische Physik der Universit\"{a}t G\"{o}ttingen,\\
Bunsenstr.9, D-37073 G\"{o}ttingen, Germany}
\author{Guillermo E. Castilla}
\address{Brookhaven National Laboratory, Upton, New York 11973}
\date{February 19, 1996}
\maketitle
\begin{abstract}
We develop a systematic method 
to treat the effect of non-linearity in the energy dispersion on the
usual bosonization result for the
single-particle Green's function of fermions
in arbitrary dimension.
The leading corrections due to the
quadratic term in the energy dispersion 
are explicitly calculated.
In the Chern-Simons theory for half-filled
quantum Hall systems curvature is shown to be essential.

\end{abstract}
\narrowtext

\begin{figure}
\epsfysize4.5cm 
\hspace{2cm}
\epsfbox{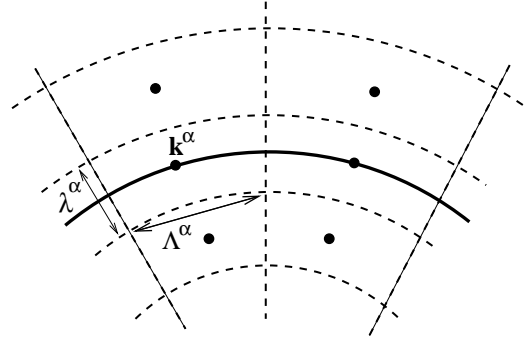}
\vspace{1cm}
\caption{
Definition of the sectors $K^{\alpha}$
and cutoffs $\Lambda^{\alpha}$ and $\lambda^{\alpha}$.
The thick solid line is the Fermi surface,
and the dots  are the origins ${\bf{k}}^{\alpha}$ of local
coordinate systems. For sectors that
intersect the Fermi surface we choose ${\bf{k}}^{\alpha}$ such that
$\epsilon_{ {\bf{k}}^{\alpha} } = \mu$, but 
in general $\epsilon_{ {\bf{k}}^{\alpha} } \neq \mu$.
By construction the union $\bigcup_{\alpha} K^{\alpha}$
covers all degrees of freedom in the system.
}
\label{fig:sectors}
\end{figure}

For many years the bosonization technique has been 
successfully used to study one-dimensional Fermi systems 
beyond perturbation theory\cite{Stone94}. 
Motivated by experimental evidence for non-Fermi liquid behavior
in high-temperature superconductors and quantum Hall systems,
the generalization of this approach to arbitrary dimensions $d$
has recently received a lot of 
attention\cite{Luther79,Haldane92,Houghton93,Kwon94,Castro94,Kopietz94,Kopietz95,Kopietzhab}.
However,
although bosonization is non-perturbative in the sense that infinitely many Feynman diagrams
are summed via an underlying Ward identity\cite{Castellani94},
it hinges on one essential approximation:
the linearization of the 
non-interacting energy  dispersion $\epsilon_{\bf{k}}$.
Namely, measuring
wave-vectors with respect to coordinate systems
centered at points ${\bf{k}}^{\alpha}$  (see Fig.\ref{fig:sectors}), 
we may write
$\epsilon_{ {\bf{k}}^{\alpha} + {\bf{q}} } =  
\epsilon_{ {\bf{k}}^{\alpha}  }  + {\bf{v}}^{\alpha} \cdot {\bf{q}} + 
\frac{ {\bf{q}}^2 }{2 m^{\alpha}}$, where ${\bf{v}}^{\alpha}$ is the
Fermi velocity and $m^{\alpha}$ is the effective mass 
close to ${\bf{k}}^{\alpha}$.
Conventional bosonization sets $\frac{1}{m^{\alpha}} = 0$.
Haldane\cite{Haldane81} has speculated 
that it should be
possible to develop a perturbation theory
around the non-perturbative bosonization solution of the linearized theory,
using $\frac{1}{ m^{\alpha}}$ as small parameter.
An attempt to construct such an expansion has recently been
made by Khveshchenko\cite{Khveshchenko95}.
However, so far his method has not been 
proven to be useful in practice. 
To the best of our knowledge,
an explicit calculation of the effect of finite $m^{\alpha}$
on the usual bosonization 
solution for the single-particle Green's function does not exist.
In the present work we shall solve this problem in arbitrary $d$ by means
of our functional bosonization approach\cite{Kopietz94,Kopietz95,Kopietzhab}.
We then use our method to study the relevance of curvature
in the Chern-Simons theory of the
half-filled Landau level\cite{Halperin93}.
This problem has recently been
examined by several authors with conflicting results\cite{Kwon94,Khveshchenko95,Altshuler94}.
We hope that our work will clarify the issue.

For simplicity let us first consider a system of spinless fermions with
Landau interaction parameters $f_{\bf{q}}^{ {\bf{k}} {\bf{k}}^{\prime}}$,
where ${\bf{q}}$ is the momentum transfer between two particles with
initial momenta ${\bf{k}}$ and ${\bf{k}}^{\prime}$.
We start by partitioning momentum space into a finite
number of sectors $K^{\alpha}$, $\alpha = 1 , \ldots , n$, 
which depend on cutoffs $\Lambda^{\alpha}$ and $\lambda^{\alpha}$ as shown in
Fig.\ref{fig:sectors}.
Assuming that the $f_{\bf{q}}^{ {\bf{k}} {\bf{k}}^{\prime}}$ 
are dominated by $ | {\bf{q}} | 
\raisebox{-0.5ex}{$\; \stackrel{<}{\sim} \;$} q_c $,
we choose $\Lambda^{\alpha} , \lambda^{\alpha} \gg q_c $.
On the other hand,
the sectors must be sufficiently small so that
the local curvature of the Fermi surface is constant within a given sector.
Furthermore, 
$f_{\bf{q}}^{ {\bf{k}} {\bf{k}}^{\prime}}$
should not change appreciably when ${\bf{k}}$ and ${\bf{k}}^{\prime}$
are restricted to given sectors, so that we may replace
$f_{\bf{q}}^{ {\bf{k}} {\bf{k}}^{\prime}} \rightarrow
f_{\bf{q}}^{ \alpha \alpha^{\prime}}$.
In particular, for Fermi surfaces with constant curvature and
interactions $f_{\bf{q}}$ that are independent of ${\bf{k}}$ and ${\bf{k}}^{\prime}$, 
{\it{there is not need any more for introducing several sectors 
as long as we can handle the curvature problem.}} In this case
we formally identify the entire momentum space
with a single sector. 
We are interested in the Matsubara Green's function
$G ( k )$, which can be represented
as a Grassmannian functional integral in the usual way\cite{Negele88,footnote1}.
We eliminate the Grassmann fields by means of a Hubbard-Stratonovich
transformation involving bosonic auxiliary fields $\phi^{\alpha}_q$ associated
with the sectors. After the standard 
transformations\cite{Kopietz94,Kopietz95,Kopietzhab}
the Green's function can be exactly written as
$G ( k ) = < [ \hat{G} ]_{kk} >_{ S_{eff} }$,
where $\hat{G}$ is an infinite matrix in momentum-- and frequency space, with
matrix elements given by 
$ [ \hat{G}^{-1} ]_{k k^{\prime}} =
 [ \hat{G}^{-1}_{0} ]_{k k^{\prime}} - [ \hat{V} ]_{k k^{\prime}}$.
Here $ [ \hat{G}_{0} ]_{ k k^{\prime} }
   =  \delta_{k k^{\prime} } G_0 (k)$,  with
$G_0 (k) =
[ i \tilde{\omega}_{n} - \epsilon_{\bf{k}} + \mu ]^{-1}$,
and the infinite matrix $\hat{V}$ is defined by
$ [ \hat{V} ]_{ k k^{\prime} }   =
 \sum_{\alpha} 
   \Theta^{\alpha} ( {\bf{k}} ) V^{\alpha}_{k- k^{\prime}}$, with
$  V^{\alpha}_{q} =
      \frac{i}{\beta} 
  {\phi}^{\alpha}_{q} $.
Here $\beta$ is the inverse temperature, $\mu$ is the chemical
potential, and the cutoff function $\Theta^{\alpha} ( {\bf{k}} )$ is unity
if ${\bf{k}} \in K^{\alpha}$, and vanishes otherwise.
Let us emphasize again that the above construction includes the case
that we identify the entire  momentum space
with a single sector:
then the $\alpha$-sums run over a single term $\alpha =1$ with
$\Theta^{\alpha} ( {\bf{k}} ) = 1$, 
and it is convenient to
choose ${\bf{k}}^{\alpha}$ such that $\epsilon_{ {\bf{k}}^{\alpha} } = \mu$.
The symbol
$< \ldots >_{S_{eff}}$ denotes
functional averaging with respect to the effective action
$ {S}_{eff}  = {S}_{2}  +  {S}_{kin}  $,
where
 ${S}_{2} = 
 \frac{{\cal{V}}}{2 \beta} \sum_{q} \sum_{\alpha \alpha^{\prime} }
  [ \underline{{f}}_{ {{q}} }^{-1} ]^{ \alpha \alpha^{\prime} }
 \phi_{-q}^{\alpha} \phi_{q}^{\alpha^{\prime}}$, and
${S}_{kin}   = 
 -  Tr \ln [ 1 - \hat{G}_{0} \hat{V} ]  $.
Here ${\cal{V}}$ is the volume of the system and $\underline{{f}}_{ {{q}} }$ is a matrix in the 
sector indices, with elements 
$ [ \underline{{f}}_{ {{q}} }]^{\alpha \alpha^{\prime}}
= f_{\bf{q}}^{\alpha \alpha^{\prime}}$. 
For $q_c \ll \Lambda^{\alpha}, \lambda^{\alpha}$ the matrix $\hat{G}$ is approximately block-diagonal,
with blocks labelled by the sector index $\alpha$.
Shifting ${\bf{k}} = {\bf{k}}^{\alpha} + {\bf{q}}$
and choosing $|{\bf{q}} | \ll \Lambda^{\alpha} , \lambda^{\alpha}$,
we have\cite{footnote1} $G ( {\bf{k}}^{\alpha} + {\bf{q}} , 
i \tilde{\omega}_n ) = < [ \hat{G}^{\alpha}]_{ \tilde{q} \tilde{q} } ] >_{S_{eff}}$,
where the Fourier transform
 ${\cal{G}}^{\alpha} ( {\bf{r}} , {\bf{r}}^{\prime} , \tau , \tau^{\prime} )$ 
of $ [\hat{G}^{\alpha}]_{\tilde{q} \tilde{q}^{\prime}}$ satisfies
 \begin{eqnarray}
  \left[- {\partial}_{ \tau} -    
  {\epsilon}^{\alpha} ( {\bf{P}}_{\bf{r}} )  + \mu - 
  V^{\alpha}  ( {\bf{r}} , \tau )
  \right]
 {\cal{G}}^{\alpha} ( {\bf{r}} , {\bf{r}}^{\prime} , \tau , \tau^{\prime} )
 &  &
 \nonumber
 \\
     =
 \delta ( {\bf{r}} - {\bf{r}}^{\prime} ) \delta^{\ast} ( \tau - \tau^{\prime} )
 & &
 \; \; \; .
 \label{eq:Galphadifrt}
 \end{eqnarray}
Here ${\bf{P}}_{\bf{r}} = -i \nabla_{\bf{r}}$ is the momentum operator
(we use units where $\hbar = 1)$,
$\epsilon^{\alpha} ( {\bf{q}} ) \equiv \epsilon_{ {\bf{k}}^{\alpha} + {\bf{q}} }$,
and $ \delta^{\ast} ( \tau )
 = \frac{1}{\beta} \sum_{n} e^{ - i \tilde{\omega}_{n}  \tau }$
 is the antiperiodic $\delta$-function.
The potential
$ V^{\alpha}  ( {\bf{r}} , \tau ) \equiv \sum_q e^{i [ {\bf{q}} \cdot {\bf{r}}
- \omega_m \tau ] } V^{\alpha}_q$
is the Fourier transform of the Hubbard-Stratonovich field
$V^{\alpha}_q = \frac{i}{\beta} \phi^{\alpha}_q$.
Eq.(\ref{eq:Galphadifrt}) together with the
boundary condition that 
 ${\cal{G}}^{\alpha} ( {\bf{r}} , {\bf{r}}^{\prime} , \tau , \tau^{\prime} )$
should be antiperiodic in $\tau$ and $\tau^{\prime}$
uniquely determines ${\cal{G}}^{\alpha}$.
Hence, to calculate the Green's function of the many-body
system, we first need to solve Eq.(\ref{eq:Galphadifrt}) for fixed
$V^{\alpha} ( {\bf{r}}, \tau )$, and then
average the result with respect to the 
effective action $S_{eff}$. Note that for $\frac{1}{m^{\alpha}} \neq 0$ the action $S_{eff}$ 
is not Gaussian. The leading non-Gaussian corrections
can be easily calculated\cite{Kopietz95,Kopietzhab}.

Solving Eq.(\ref{eq:Galphadifrt}) for finite $m^{\alpha}$ is
more difficult.
Let us make the ansatz
 \begin{equation}
 {\cal{G}}^{\alpha} ( {\bf{r}} , {\bf{r}}^{\prime} , \tau , \tau^{\prime} )
  = 
 {\cal{G}}^{\alpha}_1 ( {\bf{r}} , {\bf{r}}^{\prime} , \tau , \tau^{\prime} )
 e^{ \Phi^{\alpha} ( {\bf{r}} , \tau ) - \Phi^{\alpha} ( {\bf{r}}^{\prime} , \tau^{\prime} ) }
 \label{eq:Ansatz}
 \; \; \; .
 \end{equation}
To satisfy the boundary conditions, we require that
$\Phi^{\alpha} ( {\bf{r}} , \tau )$ should be  periodic in $\tau$, while
${\cal{{G}}}^{\alpha}_1 ( {\bf{r}} , {\bf{r}}^{\prime} , \tau , \tau^{\prime} )$ 
should be antiperiodic in $\tau$ and $\tau^{\prime}$.
The crucial observation is now that
we obtain an {\it{exact}} solution of 
Eq.(\ref{eq:Galphadifrt}) by choosing
$\Phi^{\alpha}$ and
 ${\cal{G}}^{\alpha}_1$ such that
  \begin{equation}
  \left[ - {\partial}_{ \tau} -    
  {\xi}^{\alpha} (  {\bf{P}}_{\bf{r}} ) 
  \right]
  \Phi^{\alpha} ( {\bf{r}} , \tau ) = 
   {V}^{\alpha}  ( {\bf{r}} , \tau )
   + \frac{[ {\bf{P}}_{\bf{r}} 
  \Phi^{\alpha} ( {\bf{r}} , \tau ) ]^2 }{ 2 m^{\alpha}}
  \label{eq:difPhi}
  \; \; \; ,
  \end{equation}
  \begin{eqnarray}
  \left[ - {\partial}_{ \tau} -    
  {\epsilon}^{\alpha} ( {\bf{P}}_{\bf{r}} )  + \mu
  - {\bf{u}}^{\alpha} ({  {\bf{r}} , \tau } )\cdot {\bf{P}}_{\bf{r}} 
  \right]
 {\cal{G}}^{\alpha}_1 ( {\bf{r}} , {\bf{r}}^{\prime} , \tau , \tau^{\prime} ) 
 & &
 \nonumber
 \\
 = \delta ( {\bf{r}} - {\bf{r}}^{\prime} ) \delta^{\ast} ( \tau - \tau^{\prime} )
 & &
  \; \; \; ,
  \label{eq:difG1}
  \end{eqnarray}
where $\xi^{\alpha} ( {\bf{q}} ) = \epsilon^{\alpha} ( {\bf{q} } ) - \epsilon^{\alpha} ( 0 )$, and
$ {\bf{u}}^{\alpha}({  {\bf{r}} , \tau })
 =  \frac{ {\bf{P}}_{\bf{r}}  \Phi^{\alpha} ( {\bf{r}} , \tau ) }
 { m^{\alpha}}$.
Differential equations of the type (\ref{eq:difPhi}) are called
{\it{eikonal equations}}, and appear in many fields
of physics, such as geometrical optics, 
quantum mechanical scattering theory, 
and relativistic quantum field theories\cite{Fradkin66}.
In the limit $\frac{1}{m^{\alpha}} \rightarrow 0$ 
the eikonal equation (\ref{eq:difPhi}) is
linear and can be solved exactly via Fourier transformation. The solution
has first been discussed by Schwinger\cite{Schwinger62},
see also Refs.\cite{Kopietz94,Kopietz95,Kopietzhab}.
Furthermore, in this case
the velocity ${\bf{u}}^{\alpha}({  {\bf{r}} , \tau })$  
vanishes, so that 
 ${\cal{G}}^{\alpha}_1 ( {\bf{r}} , {\bf{r}}^{\prime} , \tau , \tau^{\prime} ) 
 = 
 {{G}}^{\alpha}_{0} ( {\bf{r}} - {\bf{r}}^{\prime} , \tau - \tau^{\prime} )$,
where the Fourier transform of $G^{\alpha}_0 ( {\bf{r}} , \tau )$ is given
by $G_0^{\alpha} ( \tilde{q} ) \equiv
G_0 ( {\bf{k}}^{\alpha} + {\bf{q}} , i \tilde{\omega}_n )$.
For finite $m^{\alpha}$, Eq.(\ref{eq:difG1}) 
describes the motion of a fermion under the influence of a space-- and 
time-dependent {\it{random velocity}}
 ${\bf{u}}^{\alpha}({  {\bf{r}} , \tau })$. 
At the first sight it seems that this problem is
just as difficult to solve as the original 
Eq.(\ref{eq:Galphadifrt}). However, 
perturbation theory in terms of the 
{\it{derivative potential}}
  $ {\bf{u}}^{\alpha} ({  {\bf{r}} , \tau } )\cdot {\bf{P}}_{\bf{r}} $
in  Eq.(\ref{eq:difG1}) is {\it{less infrared singular}}
than perturbation theory in terms of the original
random potential $V^{\alpha} ( {\bf{r}} , \tau  )$
in Eq.(\ref{eq:Galphadifrt}).
Moreover, for large effective mass $m^{\alpha}$
the random velocity ${\bf{u}}^{\alpha} ( {\bf{r}} , \tau )$ is 
a small parameter which formally justifies the perturbative
treatment of the derivative potential.

Let us first consider the eikonal equation (\ref{eq:difPhi}).
Although it is impossible to solve 
this non-linear partial differential equation exactly, 
we can obtain  
the solution as series in powers of $V^{\alpha}$\cite{Fradkin66}. 
At this point it is convenient
to work in Fourier space.
Defining $\Phi^{\alpha} ( {\bf{r}} , \tau )
= \sum_{q} e^{i ( {\bf{q}} \cdot {\bf{r}} - {\omega}_m \tau  )} \Phi^{\alpha}_q$
and
$\Psi^{\alpha}_q = [ i \omega_m - \xi^{\alpha} ( {\bf{q}} ) ]
\Phi^{\alpha}_q$,
the eikonal can be written as 
 \begin{equation}
 \Phi^{\alpha} ( {\bf{r}} , \tau ) - \Phi^{\alpha} ( {\bf{r}}^{\prime} , \tau^{\prime} ) 
 = \sum_q {\cal{J}}^{\alpha}_{-q} 
 ( {\bf{r}} , {\bf{r}}^{\prime} , \tau , \tau^{\prime} )
 \Psi^{\alpha}_q
 \label{eq:eikonalpara}
 \; \; \; ,
 \end{equation}
 \begin{equation}
 {\cal{J}}^{\alpha}_{-q} 
 ( {\bf{r}} , {\bf{r}}^{\prime} , \tau , \tau^{\prime} )
 = G_0^{\alpha} ( q ) 
 \left[ 
 e^{ i ( {\bf{q}} \cdot {\bf{r}} - \omega_m \tau ) } -
 e^{ i ( {\bf{q}} \cdot {\bf{r}}^{\prime} - \omega_{m} \tau^{\prime} ) } 
 \right]
 \label{eq:Jdef}
 \; \; \; ,
 \end{equation}
where the functional $\Psi^{\alpha}_q$ satisfies
for $q \neq 0$ the non-linear integral equation
 \begin{equation}
 \Psi^{\alpha}_q = V^{\alpha}_q + \sum_{ q^{\prime} q^{\prime \prime}}
 \delta_{q , q^{\prime} + q^{\prime \prime} }
 \gamma^{\alpha}_{ q^{\prime} q^{\prime \prime} }
 \Psi^{\alpha}_{q^{\prime}}
 \Psi^{\alpha}_{q^{\prime \prime}}
 \label{eq:Psiint}
 \; \; \; ,
 \end{equation}
with the kernel given by
$ \gamma^{\alpha}_{ q^{\prime} q^{\prime \prime} }
 = \frac{ {\bf{q}}^{\prime} \cdot {\bf{q}}^{\prime \prime} }{2 m^{\alpha}}
 G_{0}^{\alpha} ( q^{\prime} )
 G_{0}^{\alpha} ( q^{\prime \prime} )$.
Here
 $G_0^{\alpha} ( {q} ) =
 [ i {\omega}_{m}
 -  {\xi}^{\alpha} ( {\bf{q}}) ]^{-1}$
is a {\it{bosonic}} Matsubara Green's function with 
energy dispersion given by the excitation energy  $\xi^{\alpha} ( {\bf{q}} )$\cite{footnote2}. 
The $q =0$-term requires a special treatment. 
From the definition $\Psi^{\alpha}_q = [ i \omega_m - \xi^{\alpha} ( {\bf{q}} ) ]
\Phi^{\alpha}_q$ it is clear that $\Psi^{\alpha}_0 = 0$.
Iterating Eq.(\ref{eq:Psiint}), we obtain a
series in powers of the random potential,
$\Psi^{\alpha}_q = \sum_{n=1}^{\infty} \Psi^{\alpha}_{n,q}$, 
where for $q \neq 0$
 \begin{equation}
 \Psi^{\alpha}_{n,q} = 
 \sum_{ q_1 \ldots q_n} \delta_{q , q_1 + \ldots + q_n}
 {C}^{\alpha}_n ( q_1 \ldots  q_n )
 V^{\alpha}_{q_1} \cdots V^{\alpha}_{q_n}
 \label{eq:Psinexp}
 \; \; \; ,
 \end{equation}
with ${C}^{\alpha}_n \propto 
(1/m^{\alpha})^{n-1}$.
The first two vertices are 
${C}^{\alpha}_1 ( q_1 ) = 1$ and
${C}^{\alpha}_2 ( q_1 q_2 ) = 
\gamma^{\alpha}_{  q_1 q_2 }$. 
Having solved Eq.(\ref{eq:Psiint}) to a certain order in
$V^{\alpha}$, we know also the random velocity
${\bf{u}}^{\alpha}({  {\bf{r}} , \tau })$
in Eq.(\ref{eq:difG1}) to the same order in $V^{\alpha}$.
In Fourier space Eq.(\ref{eq:difG1}) 
is equivalent with the Dyson equation
 \begin{equation}
  {\cal{G}}^{\alpha}_1  ( 
   \tilde{q} , \tilde{q}^{\prime} )  
  =
  \delta_{ \tilde{q} , \tilde{q}^{\prime} }  
  G^{\alpha}_0 ( \tilde{q} )
  +
  G^{\alpha}_0 ( \tilde{q} )
  \sum_{\tilde{q}^{\prime \prime} }
 {D}^{\alpha}_{  \tilde{q} , \tilde{q}^{\prime \prime} } 
 {\cal{G}}^{\alpha}_1 ( \tilde{q}^{\prime \prime} , \tilde{q}^{\prime} )  
 \; \; \; ,
 \label{eq:Dsoltotal}
 \end{equation}
where the matrix elements of the derivative potential are
$ D^{\alpha}_{ \tilde{q} , \tilde{q}^{\prime}  }
 =  
 \Psi^{\alpha}_{ \tilde{q} - \tilde{q}^{\prime} }
 \lambda^{\alpha}_{ \tilde{q} , \tilde{q}^{\prime} }$, with
dimensionless vertex
 $ \lambda^{\alpha}_{ \tilde{q} , \tilde{q}^{\prime} } =
 G_{0}^{\alpha} ( \tilde{q} - \tilde{q}^{\prime} ) 
 \frac{  ( {\bf{q}} - {\bf{q}}^{\prime} ) \cdot {\bf{q}}^{\prime} }{m^{\alpha}} $.
Iteration of Eq.(\ref{eq:Dsoltotal}) generates an expansion of
${\cal{G}}^{\alpha}_1$ in powers of the  derivative potential.
We would like to emphasize that we are not simply expanding in powers of
$\frac{1}{m^{\alpha}}$. Because the Gaussian propagator of the
$V^{\alpha}$-field is proportional to the {\it{screened}} interaction
$f^{RPA , \alpha}_q$ within random-phase approximation 
(RPA)\cite{Kopietz94,Kopietz95,Kopietzhab}, the effective
expansion parameter is proportional to
$f^{RPA , \alpha}_q / m^{\alpha}$.
This will become obvious shortly.

To obtain the Green's function
of the many-body system, we need to
average Eq.(\ref{eq:Ansatz})
with respect to the effective  action $S_{eff}$.
Because averaging restores translational invariance, we may set
${\bf{r}}^{\prime} = \tau^{\prime} = 0$ and calculate
$ {{G}}^{\alpha} ( {\bf{r}} , \tau ) =
 \left< {\cal{G}}^{\alpha} ( {\bf{r}} , 0 , \tau , 0 )
 \right>_{S_{eff}}$.
We parametrize the average Green's function as
$ {{G}}^{\alpha} ( {\bf{r}} , \tau ) 
 = 
 [ {G}^{\alpha}_1 ( {\bf{r}} , \tau )  +
 {G}^{\alpha}_2 ( {\bf{r}} , \tau ) ] e^{Q^{\alpha} ( {\bf{r}} , \tau )  }$, 
where 
 $ {G}^{\alpha}_1 ( {\bf{r}} , \tau )  
 = \left< 
  {\cal{G}}^{\alpha}_1 ( {\bf{r}} , 0 , \tau , 0) \right>_{S_{eff}}$,
 \begin{equation}
 Q^{\alpha} ( {\bf{r}} , \tau )   
 = \ln 
 \left< 
 e^{ \Phi^{\alpha} ( {\bf{r}} , \tau ) - \Phi^{\alpha} (0,0) } 
 \right>_{S_{eff}}
 \label{eq:Qdef}
 \; \; \; ,
 \end{equation}
 \begin{equation}
 {G}^{\alpha}_2 ( {\bf{r}} , \tau )  
 = \frac{ \left<
  \delta  {\cal{G}}^{\alpha}_1 ( {\bf{r}} , 0 , \tau , 0) 
 \delta e^{ \Phi^{\alpha} ( {\bf{r}} , \tau ) - \Phi^{\alpha} (0,0) } 
  \right>_{S_{eff}} }
 {\left< 
 e^{ \Phi^{\alpha} ( {\bf{r}} , \tau ) - \Phi^{\alpha} (0,0) } 
 \right>_{S_{eff}}
 } 
 \; \; \; .
 \label{eq:G2def}
 \end{equation}
Here $\delta  X  = X - <X>_{S_{eff}}$.
We now perform the averaging perturbatively. Note that
the fermionic degrees of freedom have been completely eliminated,
so that the perturbation theory is formulated in terms
of the bosonic field $V^{\alpha}_q$.
Following Refs.\cite{Kopietz95,Kopietzhab},
we calculate $Q^{\alpha} ( {\bf{r}} , \tau )$
via a linked cluster expansion.  
In this way we obtain an expansion 
$Q^{\alpha} ( {\bf{r}} , \tau )   =
\sum_{n=1}^{\infty} Q^{\alpha}_n ( {\bf{r}} , \tau )$, with
 \begin{eqnarray}
 Q^{\alpha}_n ( {\bf{r}} , \tau  ) & = &
 \sum_{q q_1 \ldots q_n} \delta_{q , q_1 + \ldots + q_n}
 W_n^{\alpha} (  q_1 \ldots q_n ) 
 \nonumber
 \\
 & \times &
 {\cal{J}}^{\alpha}_{-q}  ( {\bf{r}} ,  \tau  )
 {\cal{J}}^{\alpha}_{q_1}  ( {\bf{r}} ,  \tau  )
 \cdots
 {\cal{J}}^{\alpha}_{q_n}  ( {\bf{r}} , \tau  )
 \; \; \; ,
 \label{eq:Qngeneral}
 \end{eqnarray}
where 
 ${\cal{J}}^{\alpha}_{q_i}  ( {\bf{r}} , \tau  )
 =
 {\cal{J}}^{\alpha}_{q_i} ( {\bf{r}} , 0 , \tau , 0 )$, 
and the vertices 
 $W_n^{\alpha}$ 
can be calculated perturbatively in powers of 
our small parameter $f^{RPA, \alpha}  / m^{\alpha}$. 
At the leading tree-level (where bosonic loops are neglected) 
the vertex $W^{\alpha}_n$ 
is proportional to $(f^{RPA, \alpha})^n/(m^{\alpha} )^{n-1}$.
Already the first term $Q^{\alpha}_1 ( {\bf{r}} , \tau )$ contains non-trivial
effects due to the finiteness of $\frac{1}{m^{\alpha}}$. 
Hence, to study the relevance of curvature, 
it is sufficient to calculate $Q_1^{\alpha} ( {\bf{r}} , \tau )$
at tree-level, in which case
$W^{\alpha}_1$ is approximated by
$ - \frac{1}{2 \beta {\cal{V}} } f^{RPA, \alpha}_{q_1}$. 
This amounts to averaging in Eq.(\ref{eq:Qdef})
with respect to
$S_{eff}$ in Gaussian approximation\cite{Kopietz94,Kopietz95,Kopietzhab}, and yields
 $Q_1^{\alpha} ( {\bf{r}}  , \tau ) = R^{\alpha}_1 -
 S_1^{\alpha} ( {\bf{r}}  , \tau )$, 
with $R^{\alpha}_1 = S^{\alpha}_1 (0,0)$ and
 \begin{equation}
 S_1^{\alpha} ( {\bf{r}}  , \tau )
 = \frac{1}{\beta {\cal{V}}} \sum_q
 \frac{ f^{RPA,\alpha}_q
  \cos ( {\bf{q}} \cdot {\bf{r}} - \omega_m \tau )  }
{ [ i \omega_m - \xi^{\alpha} ( {\bf{q}} ) ][ i \omega_m + \xi^{\alpha} (
- {\bf{q}}) ] }
 \label{eq:Q1res}
 \; \; \; .
 \end{equation}
Note that $\xi^{\alpha} ( - {\bf{q}} ) =
 - \xi^{\alpha} ( {\bf{q}} ) + \frac{ {\bf{q}}^2}{ m^{\alpha}}$, 
so that for finite $m^{\alpha}$ the integrand in
Eq.(\ref{eq:Q1res}) has only simple poles.
In contrast, for $ \frac{1}{m^{\alpha}} = 0$
the denominator in Eq.(\ref{eq:Q1res}) gives rise to a
{\it{double pole}}, which
leads to rather peculiar features in the
analytic structure of the Green's function in $d > 1$
\cite{Kopietzhab,Metznerpriv}.
It is also not difficult to calculate 
the next order in $f^{RPA, \alpha} / m^{\alpha}$\cite{Kopietzhab}. 
Then one should 
retain the vertex $ W_2^{\alpha} (  q_1 q_2 ) $ at tree-level
(in which case it is approximated by $ 
\frac{1}{ ( \beta {\cal{V}} )^2} \gamma^{\alpha}_{q_1 q_2} f^{RPA, \alpha}_{q_1}
f^{RPA, \alpha}_{q_2}$)
and include one-loop corrections to $W_1^{\alpha} (  q_1 )$
(which lead to a small renormalization of  the RPA-interaction in
Eq.(\ref{eq:Q1res})). 

Next, consider the calculation of ${G}^{\alpha}_1 $ 
and $G_2^{\alpha}$.
Because the derivative potential removes possible infrared singularities,
we may use the conventional impurity diagram technique\cite{Abrikosov63}.
For the average $G^{\alpha}_1  = < {\cal{G}}^{\alpha}_1 >$
we calculate the irreducible self-energy in self-consistent
Born approximation. 
By expanding the self-energy
(and not directly the Green's function)
we take into account an infinite number of terms
in the iteration of the Dyson equation (\ref{eq:Dsoltotal}).
A truncation at a finite order would lead to unphysical multiple poles in 
the Fourier transform
${G}^{\alpha}_1 ( \tilde{q} )$ of
 ${G}^{\alpha}_1 ( {\bf{r}} , \tau ) $.
We obtain
$[{G}^{\alpha}_1 ( \tilde{q} ) ]^{-1} =  
[{G}^{\alpha}_0 ( \tilde{q} ) ]^{-1} -
{\Sigma}^{\alpha}_1 ( \tilde{q} )$,
where, to leading order in $f^{RPA , \alpha}_q / m^{\alpha}$, 
 \begin{eqnarray}
 {\Sigma}^{\alpha}_1 ( \tilde{q} )
 & = & 
  \frac{1}{\beta {\cal{V}}} \sum_{q^{\prime}}
 f^{RPA,\alpha}_{q^{\prime}}
 G_0^{\alpha} ( q^{\prime} )
 G_0^{\alpha} ( - q^{\prime} )
 \frac{ {\bf{q}} \cdot {\bf{q}}^{\prime} }{  m^{\alpha}} 
 \times
 \nonumber
 \\
 &  & \hspace{-12mm}
 \left\{
 \frac{ {\bf{q}} \cdot {\bf{q}}^{\prime} }{  m^{\alpha}} 
 G^{\alpha}_1 ( \tilde{q} + q^{\prime} )
 +
 \frac{ {{\bf{q}}^{\prime}}^2 }{ 2  m^{\alpha}}
 \left[ 
 G_1^{\alpha} ( \tilde{q} + q^{\prime} )
 - G_1^{\alpha} ( \tilde{q} - q^{\prime} ) \right]
 \right\}
 \;  .
 \nonumber
 \\
 & &
 \label{eq:sigma1res}
 \end{eqnarray}
Similarly,  we use perturbation theory to  calculate
$G_2^{\alpha}$.
The leading contribution to Eq.(\ref{eq:G2def})
is of order $f^{RPA , \alpha}/ m^{\alpha}$,
and yields for the Fourier transform
$G^{\alpha}_2 ( {\tilde{q}} )= 
G^{\alpha}_1 ( {\tilde{q}} )  Y^{\alpha} ( \tilde{q} )$, with
 \begin{eqnarray}
 Y^{\alpha} ( \tilde{q} ) & = &  
 - \frac{1}{\beta {\cal{V}}} \sum_{q^{\prime}}
 f^{RPA,\alpha}_{q^{\prime}}
 G_0^{\alpha} ( q^{\prime} )
 G_0^{\alpha} ( - q^{\prime} )
 \times
 \nonumber
 \\
 &  &
 \hspace{-12mm}
 \left\{
 \frac{ {{\bf{q}}^{\prime}}^{2}}{ m^{\alpha}}
 G^{\alpha}_1 ( \tilde{q} + q^{\prime} )
 + 
 \frac{ {\bf{q}} \cdot {{\bf{q}}^{\prime} } }{ m^{\alpha}}
 \left[ 
 G_1^{\alpha} ( \tilde{q} + q^{\prime} )
 - G_1^{\alpha} ( \tilde{q} - q^{\prime} ) \right]
 \right\}
 \; .
 \nonumber
 \\
 & &
 \label{eq:Yres}
 \end{eqnarray}
Eqs.(\ref{eq:Q1res}), (\ref{eq:sigma1res}) and (\ref{eq:Yres})
are the main result of this work.
Higher order corrections 
involve at least an additional power of $f^{RPA , \alpha}_q / m^{\alpha}$.
For $\frac{1}{m^{\alpha}} = 0$ we have $G^{\alpha}_2 = 0$
and $G^{\alpha}_1 = G^{\alpha}_0$. In $d=1$ we reproduce then the
well-known bosonization solution for the Tomonaga-Luttinger model.
Furthermore, direct expansion of our
result for $G^{\alpha}  ( {\bf{r}} , \tau )$ to first order
in $f^{RPA , \alpha}_q$ {\it{exactly}} reproduces the 
so-called GW-approximation for the self-energy\cite{Hedin65}, {\it{with
non-linear energy dispersion}}.
If we had set $\Sigma_1^{\alpha} = Y^{\alpha} = 0$,
we would have obtained a discrepancy with
the GW self-energy, because
for finite $m^{\alpha}$ the exponentiation 
$e^{Q^{\alpha}}$  of the perturbation series 
is not quite correct. 
In a sense, we have exponentiated ``too much'', so that
it is necessary to introduce correction terms
in the prefactor.  However, even in $d=1$ these corrections 
can be calculated perturbatively,
because the extra powers of ${\bf{q}}^{\prime}$ in the numerator 
of Eqs.(\ref{eq:sigma1res}) and (\ref{eq:Yres})
remove the infrared divergencies.

Our functional bosonization approach can be generalized
to include transverse gauge fields\cite{Kopietzhab,Kopietz95gauge},
so that we have now a powerful non-perturbative method for 
studying the relevance of curvature in the Chern-Simons theory
for the half-filled Landau level\cite{Halperin93}.
To leading order in the relevant small parameter (see below),
we simply need to replace $f^{RPA , \alpha}_q $ in
Eqs.(\ref{eq:Q1res}), (\ref{eq:sigma1res}) and (\ref{eq:Yres})
by the corresponding propagator
$f^{CS , \alpha}_q$ of the
transverse gauge field\cite{Halperin93,Kwon94}. 
In the most important parameter regime
$| \omega_m | \ll v_F^{\ast} | {\bf{q}} | \ll v_F^{\ast} \kappa$
(here $\kappa =  e^2 m^{\ast} / \epsilon$ is the Thomas-Fermi screening wave-vector
in $d=2$, $v_F^{\ast}$ is the effective Fermi velocity,
$m^{\ast}$ is the effective mass, and $\epsilon $ is the dielectric constant)
$f^{CS , \alpha}_q$ 
can be written as
 \begin{equation}
 f^{CS , \alpha}_q = - \frac{2 \pi  }{m^{\ast}}
 \frac{  1 - 
 (  \hat{\bf{k}}^{\alpha}  \cdot \hat{\bf{q}}  )^2 }
 {  | \omega_m | / (v_F^{\ast} | {\bf{q}} | ) + 
   | {\bf{q}} | / q_c  }
 \label{eq:fcs}
 \; \; \; ,
 \end{equation}
where we have used the Coulomb gauge and assumed a circular
Fermi surface. Here
$\hat{\bf{k}}^{\alpha} = \frac{ {\bf{k}}^{\alpha}}{ | {\bf{k}}^{\alpha} | }$,
$\hat{\bf{q}} = \frac{ {\bf{q}}}{ | {\bf{q}} | }$, 
and $q_c  = (\tilde{\phi} m^{\ast} v_F^{\ast} )^2 / \kappa$, with
$\tilde{\phi} = 2$\cite{Halperin93}. 
Substituting Eq.(\ref{eq:fcs}) into Eq.(\ref{eq:Q1res}) 
and rescaling the variables\cite{Kopietz95gauge}, we find 
$Q^{\alpha}_1 ( {\bf{r}} , \tau ) =  
g F( \tilde{r}_{\|} , \tilde{r}_{\bot} , \tilde{\tau} ; g )$,
where $ g \equiv q_c / (\tilde{\phi} m^{\ast} v_F^{\ast})$ 
is a dimensionless parameter, 
$\tilde{r}_{\|} $ [$\tilde{r}_{\bot}$] is the component of the dimensionless vector $q_c {\bf{r}}$
parallel [perpendicular] to  
$\hat{\bf{k}}^{\alpha}$,  
$\tilde{\tau} = 
v_F^{\ast} q_c \tau$, 
and  $F( \tilde{r}_{\|} , \tilde{r}_{\bot} , \tilde{\tau} ; g )$
is a dimensionless function that can be explicitly written down as a
three-dimensional integral.
Linearization of the energy dispersion corresponds to 
setting $g=0$ in the integrand 
before doing the integration.
Then it is easy to show that
$F ( \tilde{r}_{\|} , 0 ,0 ; 0) \sim 
- \frac{1}{2 \pi^2} \ln | \tilde{r}_{\|} |$ 
for $|\tilde{r}_{\|}| \rightarrow \infty$,
implying an algebraic singularity in the momentum distribution\cite{Kwon94},
just like in a Luttinger liquid. 
On the other hand, evaluation of Eq.(\ref{eq:Q1res}) with non-linear energy dispersion
corresponds to keeping $g$ finite in the integrand. 
For small $g$ we find in this case
$F ( \tilde{r}_{\|} , 0 ,0 ; g ) \sim - c  g$ for
$|\tilde{r}_{\|}|  \rightarrow \infty$,
where $c > 0$ is a numerical constant.
Thus, {\it{the algebraic singularity in the momentum distribution
is an artifact of the linearization}}. 
This gives support to the arguments put forward in Ref.\cite{Altshuler94},
and seems to agree with
the experimental fact that half-filled quantum Hall systems 
have a sharp Fermi surface\cite{Willet93}. 
We have convinced ourselves that in the 
one-dimensional Tomonaga-Luttinger model 
a finite value of $\frac{1}{m^{\alpha}}$
does {\it{not}} destroy the algebraic
singularity in the momentum distribution.
Thus, the relevance of curvature is a specific
property of gauge fields.

The precise form of the Green's function predicted by
Eqs.(\ref{eq:Q1res}), (\ref{eq:sigma1res}) and (\ref{eq:Yres})
for the half-filled Landau level
will be discussed elsewhere. 
Because by construction 
the expansion of these expressions 
to first order in the interaction
exactly reproduces lowest order
perturbation theory, the Green's function will
certainly not be of the Fermi liquid type\cite{Halperin93}.
For small $g$ it can be justified to
neglect the higher-order terms
in the eikonal expansion and in the perturbative
calculation of $\Sigma_1^{\alpha}$ and $Y^{\alpha}$.
Using the estimates of Ref.\cite{Halperin93}, we
find $ g  \approx 0.6$ in the
experimentally relevant regime\cite{Willet93}.
We are thus forced to conclude that
for an accurate quantitative comparison with experiments
higher-order terms have to be retained.

We have profited from discussions with L. Bartosch, K. Sch\"{o}nhammer,
W. Metzner, and D. V. Khveshchenko.
The work of G. C. was supported by the Division of Material
Science, U.S. Department of Energy under contract No. DE-AC02-76CH00016.

%


\begin{thebibliography}{99}
\bibitem{footnote2}
We distinguish the
fermionic $G^{\alpha} ( \tilde{q} )$ 
from the bosonic
$G^{\alpha} (q )$ by the arguments $\tilde{q}$ and $q$, see also \cite{footnote1}.
\end{thebibliography}
\end{document}